\documentclass[aps,showpacs,twocolumn]{revtex4}
\usepackage{graphicx,amsfonts}

\begin{document}

\preprint{IGC--08/4--3}

\title{Towards Loop Quantum Gravity without the time gauge.}

\author{Francesco Cianfrani}
\email{francesco.cianfrani@icra.it}
\affiliation{
ICRA-International Center for Relativistic Astrophysics, Physics Department (G9), University of Roma ``Sapienza'', Piazzale Aldo Moro 5, 00185 Rome, Italy.}

\author{Giovanni Montani}
\email{montani@icra.it}
\affiliation{ICRA-International Center for Relativistic Astrophysics, Physics Department (G9), University of Roma ``Sapienza'', Piazzale Aldo Moro 5, 00185 Rome, Italy.\\
ENEA C.R. Frascati (Dipartimento F.P.N.), via Enrico Fermi 45, 00044 Frascati, Rome, Italy.\\
ICRANET C. C. Pescara, Piazzale della Repubblica, 10, 65100 Pescara, Italy.}


\begin{abstract}

The Hamiltonian formulation of the Holst action is reviewed and it is provided a solution of second-class constraints corresponding to a generic local Lorentz frame. Within this scheme the form of rotation constraints can be reduced to a Gauss-like one by a proper generalization of Ashtekar-Barbero-Immirzi connections. This result emphasizes that the Loop Quantum Gravity quantization procedure can be applied when the time-gauge condition does not stand.

\end{abstract}

\maketitle

\section{Introduction}

The only consistent procedure to quantize non-perturbatively the gravitational field is given by Loop Quantum Gravity (LQG) (see \cite{revloop} for some reviews). One of the most impressive results within this framework consists in the discreteness of geometrical operators spectra on a kinematical level\cite{discr}, which has been recently confirmed by a suitable reduction of a spin-foam model \cite{ELPR07}. In view of this discreteness it must be demonstrated that the Lorentz invariance is still preserved on a quantum level. As far as Lorentz transformations are concerned, one must distinguish Lorentz coordinate transformations, which act on the space-time metric, from transformations on the tangent space. While the invariance under the former is insured by general covariance \cite{RS01} (see also \cite{OL}), the role of the latter is controversial in LQG. This because the local Lorentz frame is fixed before quantizing by the time-gauge condition. 
      
The Hamiltonian formulation of the Holst action \cite{Ho96} in a generic Lorentz frame was given by Barros e Sa in \cite{BS01}. Such an analysis outlines the appearance of some second-class constraints and provides a solution for them, so that only first-class constraints remain. Since the Barbero-Immirzi formulation is found by fixing the time gauge, this formulation must be related by a gauge transformation to a boosted one. However, it remains to demonstrate that the boost invariance is preserved on quantum level too.

Indeed, there are examples where first-class constraints are not associated to quantum symmetries \cite{HT}. In this respect a formulation free of the time gauge could be highly non-trivial.

Furthermore, even though the boost invariance holds, the development of a quantum theory without the time gauge can give insight on how Lorentz transformations act on a discrete space structure, especially in view of introducing elementary particles as fermion matter. 

A totally covariant approach to the quantization of gravity was proposed by Alexandrov in \cite{Al}, where Gauss constraints for the Lorentz group were inferred. Nevertheless, since in general connections do not commute, the quantization of that model is still an open issue. 

In this work, we provide a solution of second-class constraints and we develop the corresponding Hamiltonian formulation. At first we investigate the implications of fixing a generic Lorentz frame by assigning time-independent velocity components $\chi_a$ with respect to spatial hypersurfaces. In \cite{BS01} only the case $\chi_a=0$ is considered in details, while choosing different gauge conditions implies solving the boost constraints with respect to momenta conjugate to $\chi_a$.

In particular, a redefinition of variables is given such that the rotation constraints have a closed algebra and they take the form of Gauss constraints. 

When $\chi_a$ depends on time, their dynamical role must be considered and the gauge fixing is no more allowed. Nevertheless, as far as we deal with a canonical formalism for $\chi_a$ (which implements boost invariance) we recognize that their conjugate momenta are constrained to vanish. Therefore the full dynamical information is contained into the dependence on SU(2) gauge connections.  

Hence, the quantization can be performed also without any gauge fixing and no modification arises with respect to the case when the time gauge holds. For instance, the action of the area operator is investigated and the independence on $\chi_a$ of the corresponding spectrum is outlined.

Within this approach no expansion in the boost parameters is performed, thus our results extend and confirm on a different level those ones in \cite{RS01}. 

The organization of the manuscript is as follows: the Hamiltonian formulation of the Holst action is reviewed in \ref{1} and in \ref{2} second-class constraints are solved. Hence, in \ref{3} the analysis of rotation constraints is performed and the gauge structure is inferred in a time-independent frame. The extension to time-dependent frames is addressed in section \ref{6}, such that the quantization without gauge fixing and the properties of the area spectrum are described in \ref{4}. In \ref{7} the comparison with Covariant Theory is outlined. Finally, brief concluding remarks follow in \ref{5}.         
 
\section{Hamiltonian formulation}\label{1}

The action of General Relativity with the Holst modification \cite{Ho96} takes the following form (in units $8\pi G=1$)
\begin{equation} 
S=\int \sqrt{-g}e^\mu_A e^\nu_B R_{\mu\nu}^{CD}(\omega^{FG}_\mu){}^\gamma\!p^{AB}_{\phantom1\phantom2CD},\label{act}
\end{equation}

$g$ being the determinant of the metric tensor $g_{\mu\nu}$ with 4-bein vectors $e^A_\mu$ and spinor connections $\omega^{AB}_\mu$, while the expressions for $R^{AB}_{\mu\nu}$ and ${}^\gamma\!p^{AB}_{\phantom1\phantom2CD}$ are
\begin{equation}
R^{AB}_{\mu\nu}=\partial_{[\mu}\omega^{AB}_{\nu]}+\omega^A_{\phantom1C[\mu}\omega^{CB}_{\nu]},\qquad{}^\gamma\!p^{AB}_{\phantom1\phantom2CD}=\delta^{AB}_{\phantom1\phantom2CD}-\frac{1}{2\gamma}\epsilon^{AB}_{\phantom1\phantom2CD}. \label{EH}
\end{equation}

Here $\gamma$ is the Immirzi parameter. 

By a Legendre transformation, conjugate momenta ${}^\gamma\!\pi_{AB}^\mu$ can be defined. 

One easily recognizes that ${}^\gamma\!\pi^t_{AB}=0$, so $\omega_t^{AB}$ do not enter into the dynamical description. 
Let us now introduce  $\pi_{AB}^i$, such that ${}^\gamma\!\pi_{AB}^i={}^\gamma\!p^{CD}_{\phantom1\phantom2AB}\pi_{CD}^i$, whose geometrical interpretation is much more clear than ${}^\gamma\!\pi_{AB}^i$, since $\pi^i_{AB}=2\sqrt{-g}e^t_{[A}e^i_{B]}$. 

The Hamiltonian density turns out to be a linear combination of the following constraints



\begin{equation}   
\left\{\begin{array}{c}H=\pi^i_{CF}\pi^{jF}_{\phantom1D}{}^\gamma\!p^{CD}_{\phantom1\phantom2AB}R^{AB}_{ij}=0 \\\\ 
H_i={}^\gamma\!p_{AB}^{\phantom1\phantom2CD}\pi^j_{CD}R^{AB}_{ij}=0 \\\\ G_{AB}=D_i\pi^i_{AB}=\partial_i\pi^i_{AB}-2\omega_{[A\phantom2i}^{\phantom1C}\pi^i_{|C|B]}=0 \\\\ C^{ij}=\epsilon^{ABCD}\pi_{AB}^{(i}\pi_{CD}^{j)}=0 \\\\ D^{ij}=\epsilon^{ABCD}\pi^k_{AF}\pi^{(iF}_{\phantom1\phantom2B}D_k\pi^{j)}_{CD}=0
\end{array}\right..
\end{equation}

In this scenario, $H$ and $H_i$ denote the super-Hamiltonian and the super-momentum, respectively, while $G_{AB}$ are Gauss constraints of the Lorentz symmetry. Other constraints are $C^{ij}$ and $D^{ij}$, with the latter coming out as secondary ones from the former.

The full set of constraints is second-class, because $\{C^{ij},D^{kl}\}$ and $\{D^{ij},D^{kl}\}$ do not vanish on the constraint hypersurfaces. Hence $C^{ij}$ and $D^{kl}$ are not associated with any gauge symmetry.  

\section{Solution of second-class constraints}\label{2}


A generic set of 4-bein vectors can be written as
\begin{eqnarray}
e^0=Ndt+\chi_a E^a_idx^i\qquad e^a=E^a_iN^idt+E^a_idx^i.
\end{eqnarray} 

It is worth noting the role of $\chi_a$ variables, which give the velocity components of the frame $\{e^a\}$ with respect to spatial hypersurfaces \cite{FG07}. The time-gauge condition consists in $\chi_a=0$ and once adopted boost degrees of freedom are frozen. 

In what follows, we will denote $\pi^i_{0b}$ with $\pi_b^i$. 

Let us now introduce the inverses of $\pi_a^i$ ($\pi_i^a\pi^i_b=\delta^a_b$), by which the 3-metric can be written as $h_{ij}=\frac{1}{\pi}T^{-1}_{ab}\pi^a_i\pi^b_j$, with $\pi$ the determinant of $\pi^a_i$ and $T^{-1}_{ab}=\eta_{ab}+\chi_a\chi_b$. 

Defining ${}^\pi\!\omega^{\phantom1b}_{a\phantom1i}=\frac{1}{\pi^{1/2}}\pi^b_l{}^3\!\nabla_i(\pi^{1/2}\pi^l_a)$ where $^3\nabla_i$ is the covariant derivative associated with $h_{ij}$, a solution to $C^{ij}$ and $D^{ij}$ is given by
\begin{equation}
\pi^i_{ab}=2\chi_{[a}\pi^i_{b]},\qquad \omega^{\phantom1b}_{a\phantom1i}={}^\pi\!\omega^{\phantom1c}_{a\phantom1i}T^{-1b}_c+\chi_a\omega^{0b}_{\phantom{12}i}+\chi^b
(\omega_{a\phantom1i}^{\phantom10}-\partial_i\chi_a).\label{scon}
\end{equation}

We want to emphasize that conditions above are solutions of second-class constraints, without any restriction on $\chi_a$, hence at this level the Lorentz frame is not fixed at all.

\section{Rotation Constraints in a fixed local Lorentz frame}\label{3}

We will denote $\Omega$ the subspace defined by conditions (\ref{scon}) and we take $\{\omega^{0a}_{\phantom1i},\pi^i_a\}$ as coordinates on it. At this level we require $\partial_t\chi_a=0$, such that no evolutionary character is given to $\chi_a$ and they can be treated as parameters. 

Since second-class constraints have been solved, the symplectic structure on $\Omega$ is non-trivial. For instance, $\omega^{0a}_{\phantom1\phantom2i}$ do not commute among each other \cite{stu}.

Within this scheme, the Lorentz frame has been fixed. In fact, $D_i\pi^i_a=\chi^bD_i\pi^i_{ab}$, which means that the constraints associated to boosts become redundant. 




Let us now introduce ``densitized'' 3-bein of the spatial metric $h_{ij}$, which are conjugate momenta in the standard LQG formulation within the time gauge \cite{revloop} and whose expression here reads as follows

\begin{equation}
\widetilde{\pi}_a^i=S_a^b\pi_b^i,\qquad S^a_b=\sqrt{1+\chi^2}\delta^a_b+\frac{1-\sqrt{1+\chi^2}}{\chi^2}\chi_a\chi_b.
\label{dtr}\end{equation}

Hence $S^a_b$ is a map from momenta $\pi_a^i$ to 3-bein vectors of the metric on spatial hypersurfaces and it actually represents the action of the boost transformations.


Starting from the rotation constraints we sum up a vanishing contribution and multiply the result times $S^a_b$, so finding  

\begin{eqnarray}G_a=\partial_i\widetilde{\pi}^i_a+\gamma\epsilon_{ab}^{\phantom{12}c}\widetilde{A}^b_i\widetilde{\pi}_c^i=0,\label{rcon}\end{eqnarray}

{\it i.e.} \emph{SU(2) Gauss constraints are inferred also without the time-gauge condition}.

Connections $\widetilde{A}^a_i$ are given by 
\begin{equation}
\widetilde{A}_i^a=S^{-1a}_b\left(A^b_i+\frac{2+\chi^2-2\sqrt{1+\chi^2}}{2\gamma\chi^2}\epsilon^{abc}\partial_i\chi_b\chi_c\right),\label{ABI}
\end{equation}

where the $A^a_i$ variables take the following expression

\begin{equation}A_i^a=(1+\chi^2)T^{ac}(\omega_{0ci}+{}^\pi\!D_i\chi_c)-\frac{1}{2\gamma}\epsilon^a_{\phantom1cd}{}^\pi\!\omega^{cf}_{\phantom1\phantom2i}T^{-1d}_{\phantom1f}.\label{acon}\end{equation}

As far as the symplectic structure in terms of new variables is concerned, it comes out to be the canonical one, {\it i.e.} $\{\widetilde{A}^a_i(t,x),\widetilde{\pi}^j_b(t,y)\}=\delta^a_b\delta^{j}_i\delta^3(x-y)$, while the others vanish. Furthermore, it can be demonstrated that on the hypersurfaces $(\ref{scon})$ a canonical transformation maps $\{\omega^{AB}_i,\pi^j_{CD}\}$ into $\{\widetilde{A}^a_i,\widetilde{\pi}^j_b\}$ when $\partial_t\chi_a=0$. 

These features enforce the interpretation of $\widetilde{A}^a_i$ as \emph{the extension of Barbero-Immirzi connections to a generic time-independent Lorentz frame}.

The same result can be obtained within the Barros e Sa framework \cite{BS01}. That formulation was based on solving second-class constraints without fixing the boost symmetry. This implies that additional dynamical variables are present, {\it i.e.} $\chi_a$ and their conjugate momenta $\pi^a$, and that the vanishing of the boost constraints $G^{boost}_a$ provides three independent conditions. However, since $\pi^a$ appear linearly into such constraints, $G^{boost}_a=0$ can be solved with respect to them. If the obtained expressions are inserted into the rotation constraints, then we get conditions equivalent to equations (\ref{rcon}). In fact variables $A^a_i$ differ from the ones of Barros e Sa, ${}^B\!A^a_i$, only by terms containing derivatives of $\chi_a$ and no dynamical variable. This confirms the results of our analysis.

\section{On the generalization to time-dependent frames}\label{6}

The results above can be extended when $\partial_t\chi_a\neq0$ by assigning a dynamical role to $\chi_a$ themselves. This can be done by adding the corresponding conjugate momenta $\pi^a$ . In fact, a canonical transformation can be defined mapping $\{\omega_i^{AB},{}^\gamma\!\pi^j_{CD}\}$ to $\{{}^B\!A^a_i,\chi_b,\pi_c^j,\pi^d\}$ if the following conditions stand
\begin{equation}
T^{-1a}_b\pi^b+\eta^{ab}\partial_i\left(\pi^i_b-\frac{1}{\gamma}\epsilon_b^{\phantom1cd}\chi_c\pi^i_d\right)-2\eta^{[a|b|}\pi^i_b\chi^{c]}{}^B\!A_{ci}=0.\label{picon}
\end{equation}

Hence these constraints must be added to describe properly the dynamics. From the comparison with Barros e Sa's paper \cite{BS01} one recognizes that such constraints coincide with the boost ones, if the rotation constraints hold. This confirms the equivalence of the two formulations also when $\partial_t\chi^a\neq0$.

Our procedure allows to infer also in this case SU(2) Gauss constraints, by virtue of a change of phase space variables to connections $\widetilde{A}^a_i$ (this feature does not stand in Barros e Sa approach). However, now the momenta $\pi^a$ must change properly in order to deal with a canonical transformation. We denote by $\widetilde{\pi}^a$ the new expression of conjugate momenta to $\chi_a$. In terms of this new set of variables, the expressions (\ref{picon}) simplify significantly, {\it i.e.}  
\begin{equation}
B_a=T^{-1a}_{\phantom1b}\widetilde{\pi}^b=0.\label{pichi} 
\end{equation}

Therefore, a sort of decoupling occurs between $\chi_a$ and $\widetilde{A}^a_i$ variables, since Lorentz-Gauss constraints (which involve a mixing of the full set of phase-space coordinates) are equivalent to two sets of constraints, each one acting on a single couple of conjugate variables only.   

Furthermore, it is worth noting that $\chi_a$ behave as the lapse function and the shift vector, since the corresponding momenta are constrained to vanish.

\section{Loop Quantum Gravity in a local Lorentz frame.}\label{4}

Summarizing the previous analysis, the action of GR with the Holst modification can be written in a generic local Lorenz frame as follows
\begin{eqnarray}
S=\int d^4x\bigg[\widetilde{\pi}^i_a\partial_t\widetilde{A}^a_i+\widetilde{\pi}^a\partial_t\chi_a-\frac{1}{\sqrt{g}g^{tt}}H+\frac{g^{ti}}{g^{tt}}H_i+\nonumber\\+({}^\gamma\!p^{cd}_{\phantom{12}AB}+2\chi^d{}^\gamma\!p^{0c}_{\phantom{12}AB})\omega^{AB}_t\epsilon^b_{\phantom1cd}S^{-1a}_bG_a+\lambda^aB_a\bigg],\label{FINALACT}
\end{eqnarray}

$\lambda^a$ being Lagrangian multipliers. 

The action (\ref{FINALACT}) can be obtained directly from (\ref{EH}) using the definitions (\ref{scon}) and (\ref{ABI}) \footnote{This direct procedure would just allow us to write $\lambda^a$ in terms of $\omega_t^{AB}$, but nothing new would be added to previous considerations.}. 

Once a canonical quantization is performed and constraints are implemented {\it \`a la} Dirac, conditions (\ref{pichi}) can be easily solved in the most natural operator ordering taking wave functional not depending on $\chi_a$ variables. 
  
Therefore, we end up with SU(2) Gauss constraints only, a part from the super-momentum and super-Hamiltonian ones (which do not depend on the local Lorentz frame, since they are Lorentz scalars), and we formally infer the same formulation as in LQG with the time gauge. 

The quantization can now be performed starting from the holonomy-flux algebra.  

In fact, since $\widetilde{A}^a_i$ are connections of the $SU(2)$ group, holonomies $h(\widetilde{A})_\alpha$ along curves $\alpha$ and momentum fluxes $\pi(S)$ across surfaces $S$ can be defined. Their algebra is the same one as in LQG with the time gauge and this confirms that the quantization can be performed as in \cite{LOST06}. Therefore, the Hilbert space turns out to be a certain completition over the space of distributional connections, whose measure is the Ashtekar-Lewandowsky one, while basis vectors are invariant spin networks \cite{ALMMT95}. 


Within this scheme Lorentz transformations are generated by the following operators
\begin{equation}
R_a=G_a+\epsilon_a^{\phantom1bc}\chi_bB_c,\quad K_a=B_a+\frac{1-\sqrt{1+\chi^2}}{\chi^2}\epsilon_a^{\phantom1bc}G_b\chi_c.
\end{equation}

These relations can be inverted and this implies that the validity of the Hamiltonian constraints is equivalent to the invariance under the action of the Lorentz group. 

Therefore, \emph{the local Lorentz symmetry, and in particular the boost one, is actually preserved on a quantum level}.
     
For instance, let us consider the area operator $A$. Given a surface $S$ and an edge $e$ having one intersection in a point $p$ (where the tangent to $e$ does not belong to $S$), the action of the operator $A(S)$ on the parallel transport along $e$ is given by
\begin{equation}
A(S)h_e(A)=\gamma\sqrt{1+\chi^2}\sqrt{O}h_e(A),\qquad O=\frac{1}{1+\chi^2}\eta^{ab}\tau_a\tau_b,\label{area}
\end{equation}

$\tau_a$ being $SU(2)$ generators. The Immirzi parameter comes from the re-definition of generators we have to perform in order to reproduce the algebra of $G_a$.

Thus \emph{the area spectrum is discrete and this discrete structure does not depend on $\chi_a$}. 


\section{Comparison with Covariant Theory}\label{7}

The results of section \ref{4} seem to conflict with the formulation of Alexandrov \cite{Al}. In his works he attempted to quantize gravity by using a Lorentz connection $A^X_i$, $X$ running on the whole set of Lorentz indexes. Lorentz covariance is manifest, so that the Gauss constraints of a Lorentz gauge theory are inferred. Furthermore, he removed second class constraints by replacing Poisson brackets with Dirac ones, while here we use (\ref{scon}), thus solving at the same time second class constraints and the boost ones for parametric $\chi_a$. Hence, the two formulations can be matched by inserting these expressions of $\pi^i_{ab}$ and $\omega^{ab}_{\phantom{12}i}$ into the Gauss constraints of the Lorentz group, such that among them only three independent conditions remain. In this respect, it can be shown that connection (\ref{acon}) can be obtained from the one introduced in section 2 of \cite{Al} (equation (B11)) by adding some terms containing derivatives of $\chi_a$. However, Alexandrov did not consider the transformation to the true SU(2) connections (\ref{ABI}). The crucial role of such variables can be traced back to the fact that $A^X_i$ can be written via a Lorentz transformation as $(0,-\gamma\widetilde{A}^a_i)$ \cite{Alp}. 


\section{Conclusions}\label{5}
It has been proposed to extend LQG to a generic Lorentz frame and the corresponding Hamiltonian formulation has been provided. We distinguished the case of a time-independent frame from that with $\partial_t\chi_a\neq0$. In the former, $\chi_a$ can be treated as parameters and a generalization of Ashetekar-Barbero-Immirzi connections (\ref{ABI}) can be given, so finding the structure of an SU(2) gauge theory. In the latter, even though the dynamical role of $\chi_a$ must be taken into account, nevertheless the full dynamical information is contained into the dependence on $\widetilde{A}^a_i$, since physical states do not depend on $\chi_a$.

In both cases, since Gauss constraints have been inferred, the quantization of gravity can be performed. We want to point out that this is an important extension of the work by Barros e Sa \cite{BS01}, who demonstrated that the standard quantization with the time gauge is well-grounded, since the corresponding gauge condition can be safely fixed. However, he did not investigate the possibility to quantize in a moving frame or without any gauge fixing. Here, as far as we know we performed the first attempt to quantize gravity within the LQG framework taking no restriction on the local Lorentz frame.

The formulation is fully consistent with the case when the time gauge holds since the geometrical meaning of canonical variables adopted in both frameworks is the same. As a confirmation of this statement, the action of the area operator has been evaluated in a generic Lorentz frame, finding that its spectrum is discrete and it does not depend on the observer. 


This quantum symmetry reflects the invariance of the classical metric tensor under Lorentz transformations. This analysis definitely clarifies how tangent space symmetries are preserved in a quantum gravity regime.

Furthermore, this result can be linked with the findings of \cite{CSUV04}, which tightly constraint any sort of violation or modification of the Lorentz symmetry arising in a quantum gravity setting.      

\section{Acknowledgment}   

We would like to thank Sergei Alexandrov and Carlo Rovelli for their valuable comments on this subject.
F. C. would like to thank ``Fondazione Angelo Della Riccia'' and University of Roma ``Sapienza'' for having supported his work.


\begin{thebibliography}{99}

\bibitem{revloop}
C. Rovelli, ``Quantum gravity'', Cambridge University Press, Cambridge, (2004), XXIII.\\
T. Thiemann, ``Modern Canonical Quantum General Relativity'', (Cambridge University Press, Cambridge, England, 2006).\\
F. Cianfrani, O.M. Lecian, G. Montani, ``Fundamentals and recent developments in non-perturbative canonical Quantum Gravity'',  submitted to {\it Rep. Prog. Phys.}.

\bibitem{discr}
C. Rovelli, L. Smolin, {\it Nucl. Phys. B}, {\bf 442}, (1995), 593-622; Erratum-ibid. {\bf 456}, (1995), 753.\\
A. Ashtekar, J. Lewandowski, {\it Class. Quant. Grav.}, {\bf 14}, (1997), A55-A82.

\bibitem{ELPR07}
J. Engle, E. Livine, R. Pereira, C. Rovelli,  {\it Nucl. Phys. B}, {\bf 799}, (2008), 136. 

\bibitem{RS01}
C. Rovelli, S. Speziale, {\it Phys. Rev. D}, {\bf 67}, (2003) 064019.

\bibitem{OL}
E. R. Livine, D. Oriti, {\it JHEP}, {\bf 0406}, (2004) 050. 

\bibitem{Ho96}
S. Holst, {\it Phys. Rev. D}, {\bf 53}, (1996) 5966-5969. 

\bibitem{BS01}
N. Barros e Sa, {\it Int. J. Mod. Phys. D}, {\bf 10}, (2001), 261-272.

\bibitem{HT}
M. Henneaux, C. Teitelboim, \emph{Quantization of Gauge Systems}, Princeton University Press, (1994).

\bibitem{Al}
S. Alexandrov, E. R. Livine, {\it Phys. Rev. D}, {\bf 67}, (2003), 044009.

\bibitem{FG07}
F. Cianfrani, G. Montani, {\it Class. Quantum Grav.}, {\bf 24}, (2007) 4161-4168. 

\bibitem{stu}
F. Cianfrani, G. Montani, `` The Role of Time-Gauge in Quantizing Gravity'', to appear on proceedings of the III Stueckelberg workshop, Pescara July 8-18 (2008), in preparation.


\bibitem{LOST06}
J. Lewandowski, A. Okolow, H. Sahlmann, T. Thiemann, {\it Comm. Math. Phys.}, {\bf 267}, No. 3,(2006) 703-733. 

\bibitem{ALMMT95}
A. Ashtekar, J. Lewandowski, D. Marolf, J. Mourao, T. Thiemann, {\it J. Math. Phys.}, {\bf 36}, (1995), 6456-6493. 

\bibitem{Alp}
S. Alexandrov, private communication.

\bibitem{CSUV04}J. Collins, A. Perez, D. Sudarsky, L. Urrutia, H. Vucetich, {\it Phys. Rev. Lett.}, {\bf 93}, (2004), 191301. 

\end{thebibliography}
\end{document}